\documentclass[twocolumn, prb,superscriptaddress]{revtex4-1}
\usepackage{hyperref}

\usepackage{fullpage}
\usepackage{graphicx}
\usepackage{xcolor}
\usepackage{braket}
\usepackage{amsmath}
\usepackage{tikz}
\usepackage{verbatim}
\usepackage{bookmark}
\usetikzlibrary{decorations.pathmorphing}
\usetikzlibrary{arrows}
\usepackage{nicefrac}

\begin{document}
\title{A comparison of computed and experimental neutron diffraction intensity at large momentum for MnO and NiO} 

\author{Alexander R. Munoz}
\affiliation{Department of Physics; University of Illinois at Urbana-Champaign, Urbana IL 61801}
\author{Lazar Kish }
\affiliation{Department of Physics; University of Illinois at Urbana-Champaign, Urbana IL 61801}
\author{Kannan Lu}
\affiliation{Department of Physics; University of Illinois at Urbana-Champaign, Urbana IL 61801}
\author{Thomas Heitmann}
\affiliation{The Missouri Research Reactor, University of Missouri, Columbia, Missouri 65211}
\author{Gregory J. MacDougall} 
\affiliation{Department of Physics; University of Illinois at Urbana-Champaign, Urbana IL 61801}
\author{Lucas K. Wagner}
\affiliation{Department of Physics; University of Illinois at Urbana-Champaign, Urbana IL 61801}
\begin{abstract}
Magnetic neutron scattering measures spin-spin correlations giving information about the long-range spin order as well as the shape of the spin density in magnetic materials.
Similarly, detailed first principles calculations directly compute the spin density in materials.
In this work, the authors carefully compare experimentally measured magnetic neutron intensities to three levels of theory: density functional theory in two approximations, and fixed-node diffusion Monte Carlo.
While each theory performs similarly for the simple antiferromagnet MnO, there are significant differences between density functional theory and diffusion Monte Carlo in NiO.
For both materials, fixed-node diffusion Monte Carlo shows the lowest RMS error with respect to experiment for the form factor and the intensity. 
Through connection of the intensities and form factors to the real-space spin densities, it is shown that diffusion Monte Carlo spin density becomes more diffuse by spreading spin away from the bond directions.
This benchmark is of importance when considering the efficacy of an \textit{ab initio} calculation in capturing both the long-range magnetic correlations and the microscopic spin details of magnetic systems.
\end{abstract}

\maketitle

\section{Introduction}
Understanding magnetism and its interactions with other degrees of freedom is a major goal in the study of materials.
As neutron sources have become more powerful, they have become capable of capturing high quality data of the equal time spin-spin correlation function $S(\textbf{q})$, often called the structure factor.
At large values of $\textbf{q}$, the structure factor gives information about the microscopic details of the spin density.
In cases such as orbital ordering\cite{orbital_order}, the cuprates\cite{tranquada2,hole_doped}, and iron pnictides\cite{iron_based}, the microscopic details of the spin density within the unit cell can offer significant insights into the behavior of the system and serve as a powerful benchmark of the capability of computational techniques to describe the spin behavior of solids.

First principles methods such as density functional theory (DFT) have been used to model magnetic behavior of solids.\cite{dft_form,schron,dft_pnictides,qureshi_dft,cluster} 
Standard density functionals with corrections, such as DFT+U\cite{lsdau_nio} or DFT+DMFT\cite{dft+dmft}, have been used to model magnetic behavior and predict magnetic form factors accurately.
Corrections are often used because standard DFT functionals treat electron correlation and interactions only approximately.

An alternative approach is fixed-node diffusion Monte Carlo (DMC) which is a fully first principles technique that is not based on a correction to density functional theory, and has been shown to accurately describe a number of materials with magnetic and charge order.\cite{dmc_nio,vo2_huihuo,wag_abba,superexchange_cuprates,dmc_feo,compton_vo2}
The density-density correlation function was recently computed in DMC for graphite and graphene and found to agree very well with inelastic x-ray experiments.\cite{sigma_huihuo}
In principle, the structure factor is also directly available in the DMC calculation, and may be more accurate than that computed using DFT functionals, which can only access the spin density and not the structure factor.
\begin{figure}
\includegraphics{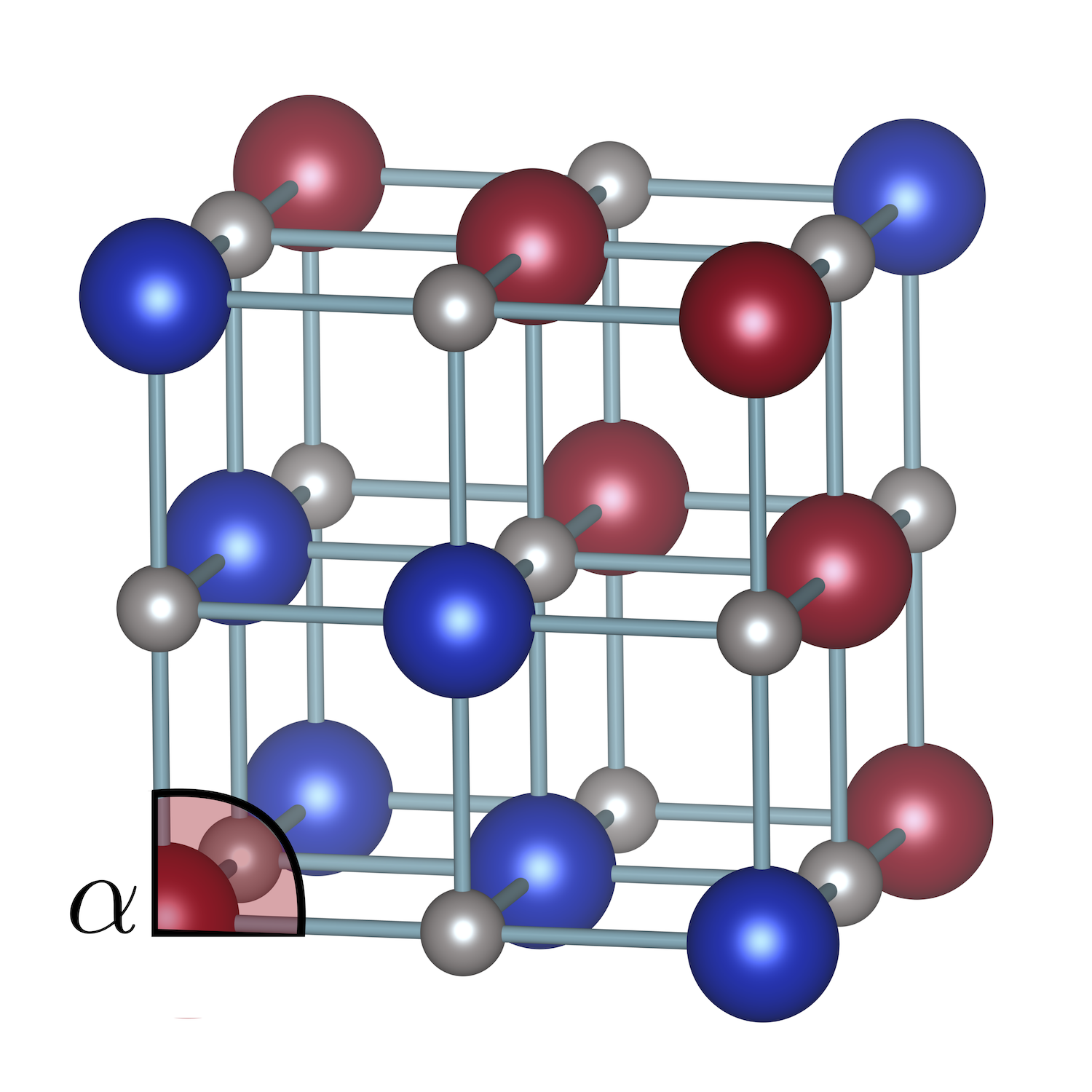}
\caption{Antiferromagnetic order in MnO and NiO. Transition metal ions are indicated by the red and blue spheres with the color indicating the spin direction on each site. The grey spheres are O sites. The angle $\alpha$ is the angle associated with the rhombohedral distortion experienced in both systems. Visualization produced with VESTA.\cite{vesta}}
\label{fig:structure}
\end{figure}

In this manuscript, we benchmark DMC and DFT calculations of the magnetic structure factor and compare directly to neutron scattering experiments. 
The structure factor is directly computed in DMC to produce magnetic neutron scattering intensities for MnO and NiO in their magnetic ground states, shown in Figure~\ref{fig:structure}.
We also compute the Fourier transform of the ion magnetization, called the magnetic form factor, in both materials to make a direct comparison between the spin densities calculated in DFT and DMC.
The neutron powder diffraction experiments were performed on both materials at the University of Missouri Research Reactor (MURR).
The results are separately normalized using sum rules and compared directly without fitting parameters.
With this comparison, we are able to benchmark the computational methods using the microscopic information inherent to the neutron intensity and form factor.

\section{Methods}
\subsection{Theory} 

We begin with the first-principles electronic Hamiltonian under the Born-Oppenheimer approximation, 
\begin{equation}
	\hat H = -\frac{1}{2}\sum_i \nabla_i^2 + \sum_{ij} \frac{1}{r_{ij}}- \sum_{\alpha, i} \frac{Z_\alpha}{r_{i\alpha}}+\sum_{i\alpha}\hat V_{\text{ecp}}(r_i-R_\alpha),
\end{equation}
where $i$ is the electron index, $R_\alpha$ is the position of an ion, $r_i$ is the position of electron $i$, $\alpha$ is an index over ions, and $V_{\text{ecp}}$ is the pseudopotential from Burkatzki, Filippi, and Dolg.\cite{bfd_1}

\begin{figure}
\includegraphics{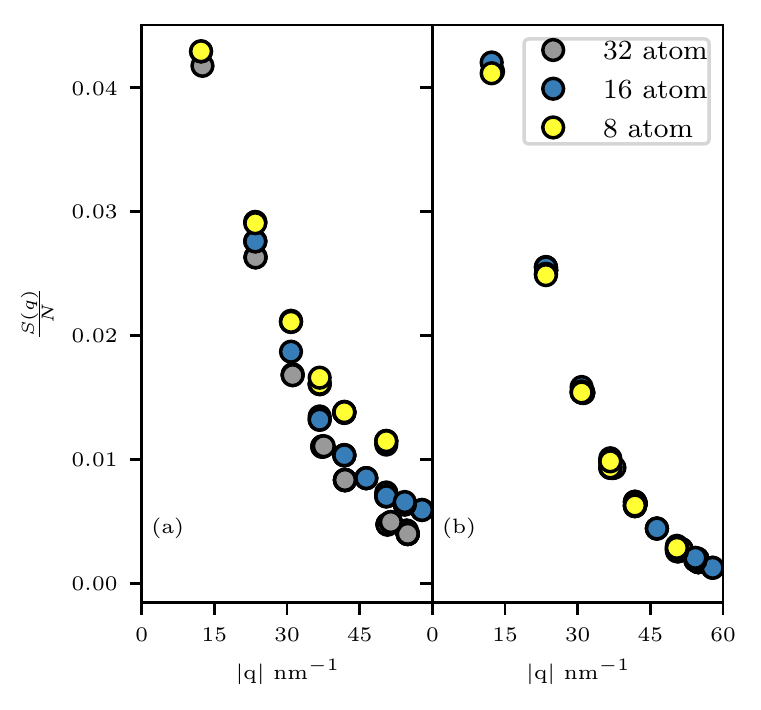}
\caption{(a): The calculated magnetic Bragg peaks for the three system sizes of MnO. (b): Calculated magnetic Bragg peaks with incoherent contributions subtracted.} 
\label{fig:finite_size}
\end{figure}

DFT calculations were performed using the CRYSTAL\cite{crystal} package with the Perdew-Burke-Ernzerhof (PBE) and PBE0 exchange-correlation functionals.
Orbitals in the materials are described using the BFD triple-$\zeta$ basis where core electrons were accounted for by the BFD pseudopotential.\cite{bfd_1,bfd_2}
For these calculations, we adopt the structural information from Schr{\"o}n with the Mn-Mn distance of 4.44 {\AA} and the Ni-Ni distance of 4.17 {\AA}. \cite{schron}
Below $T_N$, both MnO and NiO undergo rhombohedral distortions (space group $R\bar 3M$) which we denote by the angle $\alpha$ in Figure~\ref{fig:structure}.
For MnO, we use the angle 89.4 degrees and for NiO we use the angle 89.94 degrees.\cite{schron,Balagurov}
With this geometry, we begin constructing the trial wave function for DMC from the PBE0 solution for the antiferromagnetic state of both materials, shown in Figure~\ref{fig:structure}; that is, the spin broken determinant from PBE0 will serve as the Slater determinant in our trial wave function.

Fixed-node diffusion Monte Carlo (DMC) is a stochastic projection method for finding the ground-state of a Hamiltonian, $\hat H$.\cite{foulkes_rev}
DMC projects a trial wave function $|\psi_T \rangle$ towards the ground state, $|\psi_{\text{FN}} \rangle$ as follows:
\begin{equation}
	\lim_{\tau \rightarrow \infty} e^{-\tau \hat H_{\text{FN}}} |\psi_T \rangle = |\psi_\text{FN} \rangle, 
\end{equation} 
where $\tau$ is the time-step and $\hat H_{\text{FN}}$ is the fixed node Hamiltonian, which has infinite potential at the nodes of the trial wave function.
DMC inherits the nodes of the trial wave function and gives the lowest energy wave function for those nodes, while maintaining a variational upper bound.
The DMC trial wave function is the spin-symmetry broken PBE0 Slater determinant multiplied by a two-body Jastrow factor.\cite{qmc_opt,qmc_opt_2,qmc_opt_3}
It has been found that for these materials the PBE0 determinant minimizes the variational fixed node energy. \cite{pbe0,pbe0_node}
The Jastrow parameters in the trial wave function are optimized with the energy optimization introduced by Toulouse and Umrigar. \cite{qmc_opt,qmc_opt_2}

After running DMC with the optimized trial wave function, the magnetic structure factor is calculated. 
The magnetic structure factor is a measure of the spin-spin correlations of the unpaired electrons in the system\cite{tranquada,corr_neut},
\begin{equation}
	S(\textbf{q}) = \left\langle \Psi \left|  \sum_{i \neq j} s_i s_j e^{i \textbf{q} \cdot  \textbf{r}_{ij}} \right|\Psi\right\rangle ,
\end{equation}
where $s_i$ and $s_j$ are the spins of electrons taking the values $\pm 1$, $\textbf{q}$ is the scattering vector defined by reciprocal lattice vectors, and $\textbf{r}_{ij}$ is the distance between a pair of electrons.
This quantity is twist-averaged over 8 k-points in the Brillouin zone for both DMC and DFT.
\begin{table}[]
\begin{tabular}{|c|c|c|c|c|c|c|}
	   \hline
	   & $\mu_{\text{B,Exp.}}$     & $\mu_{\text{B,DMC}}$ & $\mu_{\text{B,PBE0}}$ & $\mu_{\text{B,PBE}}$ & $\alpha$ (Deg.) & a (\AA)   \\
	   \hline
MnO        &  4.55(1) & 4.59(1) & 4.62(1) & 4.25(2) & 89.4 & 2.22 \\
	   \hline
NiO        &  1.9(1) & 2.01(3) & 1.96(4) & 1.41(4) & 89.94 & 2.08 \\
	   \hline
\end{tabular}
\caption{The lattice parameters and magnetic moments for MnO and NiO. The magnetic moments for each material were determined by experiment and each level of theory. The lattice parameters were determined by Rietveld refinement where $\alpha$ is the angle indicated in Figure~\ref{fig:structure} and a is the transition metal to ligand distance.}
\end{table}

To further study the microscopic magnetic information, we calculate the magnetic form factor.
The form factor is defined as the Fourier transform over the magnetization density\cite{squires},
\begin{equation}
	f(q) = \int m(\textbf{r}) e^{i\textbf{q} \cdot \textbf{r}} d^3r,
\end{equation}
where $m(\textbf{r})$ is the magnetization of the isolated atom.
In practice, we isolated a transition metal from each solid for each technique and identified the zeros of the magnetization.
We then perform the Fourier transform of this magnetization and take the magnitude squared to compare to experiment.
\begin{figure}
\includegraphics{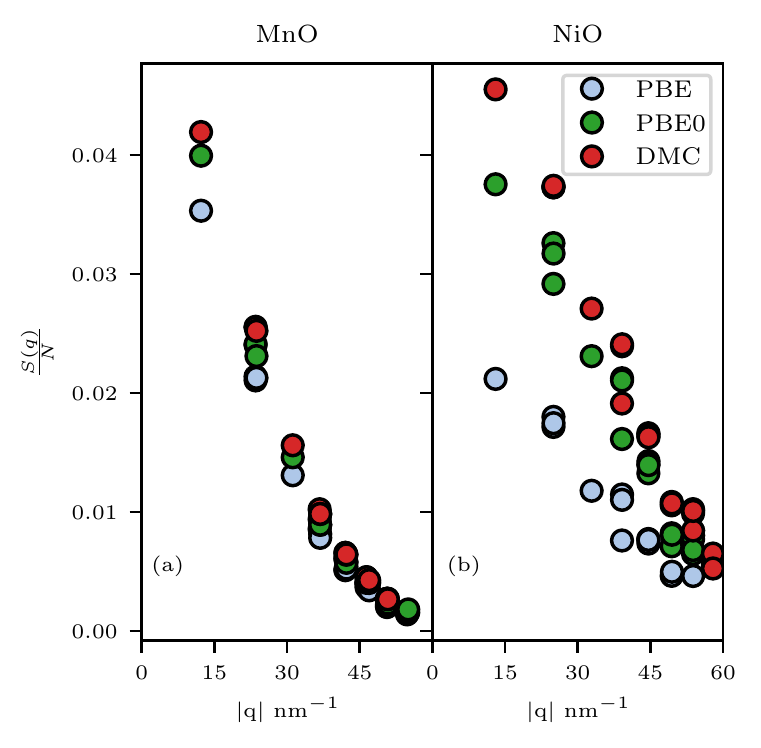}
\caption{Theory comparison between PBE, PBE0, and DMC. In both cases, DMC shows the highest value for the structure factor and PBE obtains the smallest. In the case of NiO, PBE severely underestimates the structure factor.}
\label{fig:theory_comp}
\end{figure}

\subsection{Experiment} 

Neutron powder diffraction was performed on MnO and NiO powders. 
Polycrystalline samples of MnO and NiO of 99.99 percent purity were obtained from Sigma Aldrich Inc. 
Sample quality was verified by powder x-ray diffration on a Siemens/Bruker D-5000 XRD System at the Illinois Materials Research Laboratory using Cu K-radiation at 1.5418 {\AA} at ambient temperature.
The structure determined by x-ray diffraction was consistent with previous neutron scattering studies.\cite{jacobson}\cite{Balagurov}
Impurity phases were ruled out using the software Jade.\cite{jade}
Neutron powder diffraction was performed on the high-resolution Position Sensitive Detector powder diffractometer at MURR.
The diffractometer uses a segmented silicon focusing monochromator (Si (511) reflection= 1.485 \AA) and an oscillating radial collimator to eliminate environmental scattering.
Powders were mounted in a V sample cell in a He environment in an Al exchange gas can for improved thermal contact.

Diffraction profiles were obtained for each material in both their paramagnetic and ordered phases. 
The respective measurements were made at 300K for MnO and 540K for NiO, and at 7.7K for both materials.
At the higher temperatures, both materials adopt a rocksalt structure.
Below 525 Kelvin and 116 Kelvin, NiO and MnO undergo rhombohedral distortions on the order of 0.06 degrees and 0.6 degrees, respectively.
Rietveld refinements were performed using the FullProf Suite\cite{fullprof} to obtain structural parameters and magnetic moments.
Our results were in good agreement with previous experimental studies. \cite{Balagurov}
Integrated peak intensities for form factor analysis were obtained using separate fits to Gaussian profiles.
Extraction of the atomic form factor requires normalization of the intensity by the multiplicity, polarization corrections, and the geometric Lorentz factor for powder measurements. \cite{furrer}

\subsection{Comparison between experiment and theory} 
\begin{figure}
\includegraphics{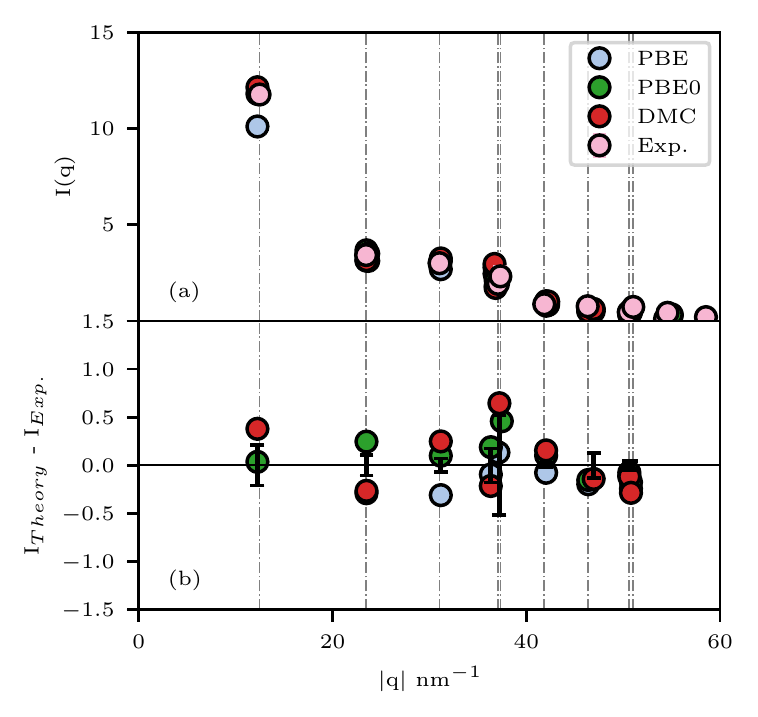}
\caption{(a): Experimental and theoretical magnetic neutron scattering intensities for MnO. (b): Theoretical values with experimental values subtracted. For high values of \textbf{q}, there are missing theoretical intensities because incoherent signals dominate the Bragg peak signal.}
\label{fig:mno_intensity}
\end{figure}

\begin{figure}
\includegraphics{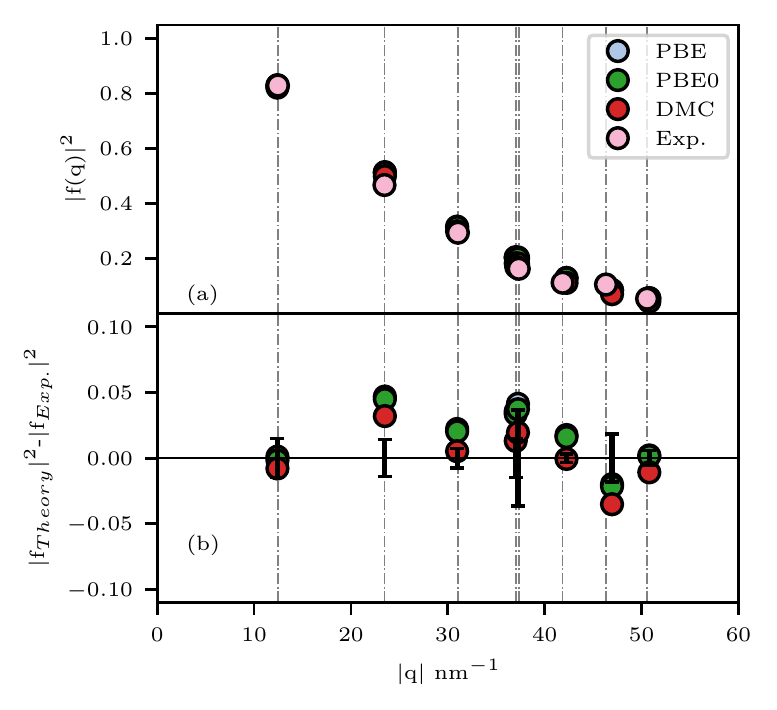}
\caption{(a): Theoretical and experimental magnetic form factors for MnO. (b) Theoretical values with experimental values subtracted. DMC (Red) has the smallest errors for \textbf{q} less than 45 nm$^{-1}$.
}
\label{fig:mno_fq}
\end{figure}

To draw direct comparisons between theory and experiment, we compare both the magnetic neutron intensity and the atomic form factor, beginning with the calculation of the magnetic neutron intensity.
Intensities are calculated by multiplying Equation 3 by the polarization factor and the multiplicity of the reflections implicit to the symmetry of the magnetic lattice.\cite{nist}
The intensity for experimental coherent magnetic scattering is given as follows,
\begin{equation}
	I(\textbf{q}) = C(1-(\hat{q} \cdot \hat{\mu})^2) S(\textbf{q})
	\label{eqn:intensity}
\end{equation}
where $C$ contains constants including the Debye-Waller factor, the term $(1- (\hat q \cdot \hat \mu)^2)$ ensures the magnetic moment direction $\hat \mu$ is perpendicular to the propagation vector, and $S(\textbf{q})$ is given by Equation 3.
We use neutron powder diffraction to determine the in-plane, (001), and out-of-plane components of the magnetic moment direction.
Theoretically, we use the components of the magnetic moment direction and average over the possible domain directions.
To compare theoretical and experimental intensities directly, we normalize the intensity using the total moment sum rule,
\begin{equation}
\int d\omega \int d\textbf{q} S(\omega,\textbf{q}) = s(s+1),
\end{equation}
where $\omega$ is the frequency, and s is the magnetic moment on the magnetic ions.
The magnetic moments obtained from experiment, DFT, and DMC are shown in Table I.
The experimental values for the size of the magnetic moment are in agreement with previous studies.\cite{schron} \cite{Balagurov}
For this comparison, there is only a single parameter, the direction of the spin moment taken from experiment, which adjusts the theoretical values according to Eqn~\ref{eqn:intensity}.

Due to the additional processing necessary to obtain the experimental form factor, we consider the intensity a better quantitative metric for our benchmark, but the form factor does offer further qualitative information.
The comparison between theory and experiment requires no adjustment as the form factor is normalized to one when \textbf{q} is equal to zero by definition. 

\section{Results}

\subsection{Theoretical results}
Finite-size effects for the structure factor $S(\textbf{q})$ were studied by performing calculations on 8-atom, 16-atom, and 32-atom cells of MnO.
Figure~\ref{fig:finite_size} shows the calculation of Equation 3 normalized by the number of electrons in each calculation cell.
This normalization is performed because the Bragg peaks scale with the number of electrons in the system.
Figure~\ref{fig:finite_size}a shows the magnetic Bragg peaks for each calculation cell size.
Removing the incoherent contributions from the magnetic Bragg peaks gives Figure~\ref{fig:finite_size}b where it is clear that DMC has insignificant finite-size effects.
Quantitatively, the largest and second largest systems were within 5 percent of one another.
From this point, we will carry on using the 16-atom calculation for further study.
\begin{figure}
\includegraphics{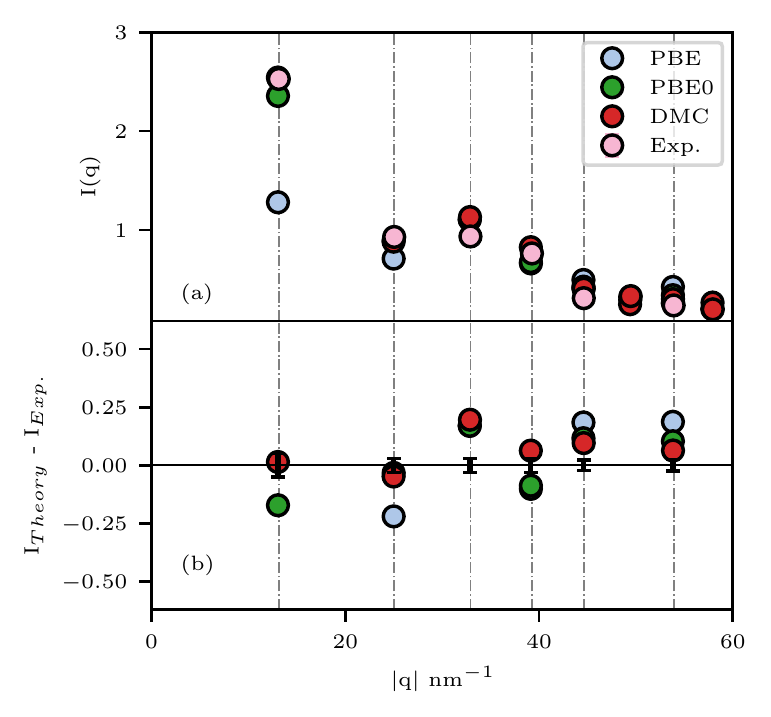}
\caption{(a): Theoretical and experimental magnetic neutron scattering intensities for NiO between theory and experiment. The NiO intensity shows a discrepancy at $\sim$32 nm$^{-1}$, the Ni to O distance. (b): Theoretical values with experimental values subtracted. There are missing experimental Bragg peaks at $\sim$49 nm$^{-1}$ and $\sim$57 nm$^{-1}$ due to the overlaps of the structural and magnetic peaks. The first PBE Bragg peak is off the scale of the difference plot.
}
\label{fig:nio_intensity}
\end{figure}

Figure~\ref{fig:theory_comp} compares $S(\textbf{q})$ at the magnetic Bragg peaks as calculated with PBE, PBE0, and DMC, for both MnO and NiO.
The theoretical results are in much better agreement for MnO than for NiO, due to the d$^5$ nature of MnO leading to a spherical spin distribution. 
PBE underestimates the magnetic moment of NiO significantly, leading to a reduced scale of $S(\textbf{q})$ overall. 
We will separate the overall structure factor from the shape of the curve when comparing to experiment by examining the form factor. 

Figure~\ref{fig:theory_comp}b shows the calculated structure factor for NiO.
Since NiO is anisotropic, there is some splitting of the Bragg peaks, most notably at 39 nm$^{-1}$. 
This was not observed experimentally due to the resolution of the instrument. 
By comparing the size of the splitting, we see that DMC and PBE0 result in larger anisotropy than PBE, likely indicating an error in the spin density of PBE. 
\begin{figure}
\includegraphics{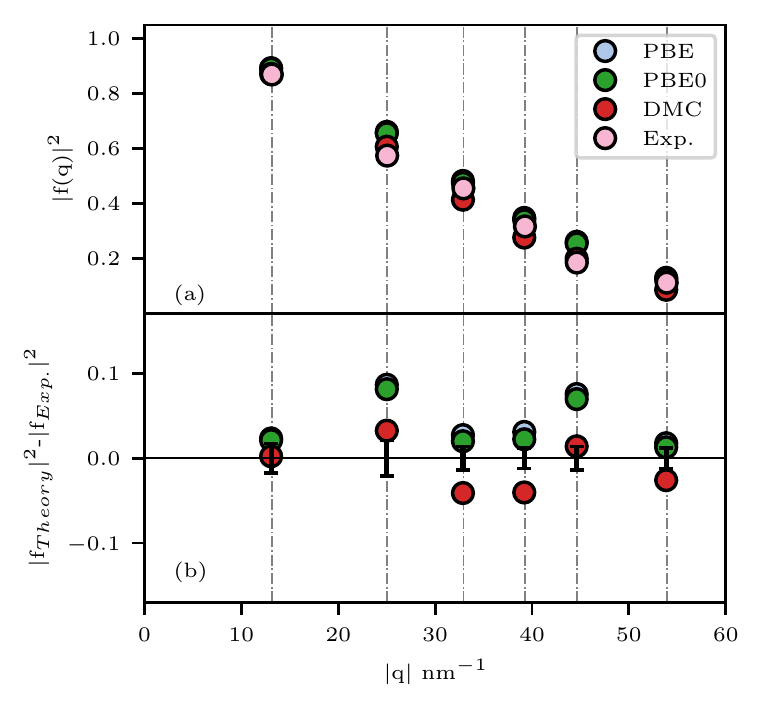}
\caption{(a): Theoretical and experimental magnetic form factors for NiO. (b): Differences between theoretical and experimental form factors. PBE (Blue) and PBE0 (Green) overestimate the form factor at each peak. DMC (Red) is the closest to experiment while underestimating the form factor.}
\label{fig:nio_fq}
\end{figure}

\subsection{Comparison between experiment and theory}
\subsubsection*{MnO}

The normalization between the levels of theory and experiment requires the magnitude and the direction of the magnetic moment.
The magnitude of the magnetic moment is simple to compute, shown in Table I.
For MnO, the magnitudes of the magnetic moment for DMC, PBE0 and experiment are in agreement with previous studies.\cite{schron} \cite{Balagurov} 
Our experiment indicates that the magnetic moment in MnO is in the (001)-plane.
The direction of the magnetic moment is used in the calculation of the intensity for each level of theory and the magnitude of the magnetic moment sets the normalization of the intensity.

The intensities from experiment and each level of theory for MnO are shown in Figure~\ref{fig:mno_intensity}.
In Figure~\ref{fig:mno_intensity}a, we see that each peak of the theoretical intensities follows the trend of the experimental results, but there are missing theoretical Bragg peaks for high values of \textbf{q} because the incoherent signal begins to dominate the Bragg peaks.
From the intensity, each level of theory calculates the structure factor above 40 nm$^{-1}$ with the same accuracy.
Figure~\ref{fig:mno_intensity}b shows the difference between the theoretical and experimental intensities.
For the first Bragg peak at 13 nm$^{-1}$, PBE underestimates the intensity. 
This distance in real-space corresponds to the Mn to Mn distance where we can expect PBE to underestimate the structure factor given that it underestimates the magnetic moment.

\begin{table}[]
\begin{tabular}{|c|c|c|c|}
	   \hline
	   & PBE     & PBE0 & DMC      \\
	   \hline
I(q)       & 0.59(3) & 0.21(3) & 0.31(6)  \\
	   \hline
$|$f(q)$|$$^2$ & 0.0283(3)  & 0.0266(2) & 0.0194(1)  \\
           \hline
\end{tabular}
\caption{RMS errors of intensity and form factor with respect to experiment for MnO}
\end{table}

The theoretical and experimental atomic form factors for MnO are shown in Figure~\ref{fig:mno_fq}a.
Here, the theoretical shape of the spin density agrees with the experiment within estimated uncertainties. 
The form factor shows DMC spreads its spin density farther away from the center of the Mn ion than either of the other two methods.

To summarize the overall agreement between theory and experiment, we use the root mean square error (RMSE) with respect to the experiment, shown in Table II.
For the MnO intensity, we see DMC and PBE0 performing similarly while PBE has a larger RMSE due to its consistent underestimate of the magnetic moment, as shown in Table I. 

\subsubsection*{NiO}
For NiO, the calculated magnetic moments are shown in Table I where PBE0 and DMC are in agreement with our experimental magnetic moment.
In NiO, PBE has a much larger percentage error in the magnetic moment.
The experimental magnetic moment for NiO has 1.853 $\mu_B$ in the (001)-plane and 0.589 $\mu_B$ in the z-direction giving a net magnetic moment of 1.9(1) $\mu_B$, in agreement with Schr{\"o}n.\cite{schron}

\begin{table}[]
\begin{tabular}{|c|c|c|c|}
	   \hline
	   & PBE     & PBE0 & DMC      \\
	   \hline
I(q)       & 0.53(3) & 0.123(2) & 0.098(1)  \\
	   \hline
$|$f(q)$|$$^2$ & 0.0511(4)  & 0.0464(4) & 0.0296(1)  \\
           \hline
\end{tabular}
\caption{RMS errors of intensity and form factor with respect to experiment for NiO}
\end{table}

The intensity for NiO is shown in Figure~\ref{fig:nio_intensity}.
Due to the underestimate of the magnetic moment in PBE, it underestimates the intensity at the first Bragg peak by approximately half of the value, resulting in a large disagreement with experiment. 
For all three methods, the intensity has a discrepancy from experiment at $\sim$32 nm$^{-1}$, which corresponds to the Ni to O bond distance in real space.
Since the magnetic moment on the Ni sites is shown to agree with experiment and the principal Bragg peak overlaps between experiment and DMC, we suspect that the additional intensity at $\sim$32 nm$^{-1}$ is due to excess spin density on the O sites. 

Across the range of \textbf{q}, we see in Figure~\ref{fig:nio_intensity}b that DMC has the lowest difference with respect to experiment.
In contrast to MnO, PBE0 underestimates the first Bragg peak despite calculating the correct magnetic moment. 
PBE, on the other hand, fails to capture the trend of the intensity.
The RMSE for the NiO intensity, shown in Table III, shows the performance between the theories.  

The calculated form factor for NiO is shown in Figure~\ref{fig:nio_fq}a.
The PBE and PBE0 form factors are overestimated across several values of \textbf{q} and are consistently higher than DMC, as shown in ~\ref{fig:nio_fq}b. 
In real space, \textbf{q} values above 40 nm$^{-1}$ correspond to the on-site spin distribution of the Ni atoms.
While DMC maintains the trend in peak values, it corrects the PBE0 starting point toward the experimental form factor.
The atomic form factor also agrees with experiment at 32 nm$^{-1}$ further indicating that the intensity overestimate at 32 nm$^{-1}$ is due to an overestimate of the O spin.
The RMSE, shown in Table III, shows DMC more accurately calculates the NiO form factor. 
In order to understand how DMC is differentiating itself from DFT, we turn to the real space spin density to show how DMC organizes the spin on the Ni site.
\begin{figure*}
\includegraphics{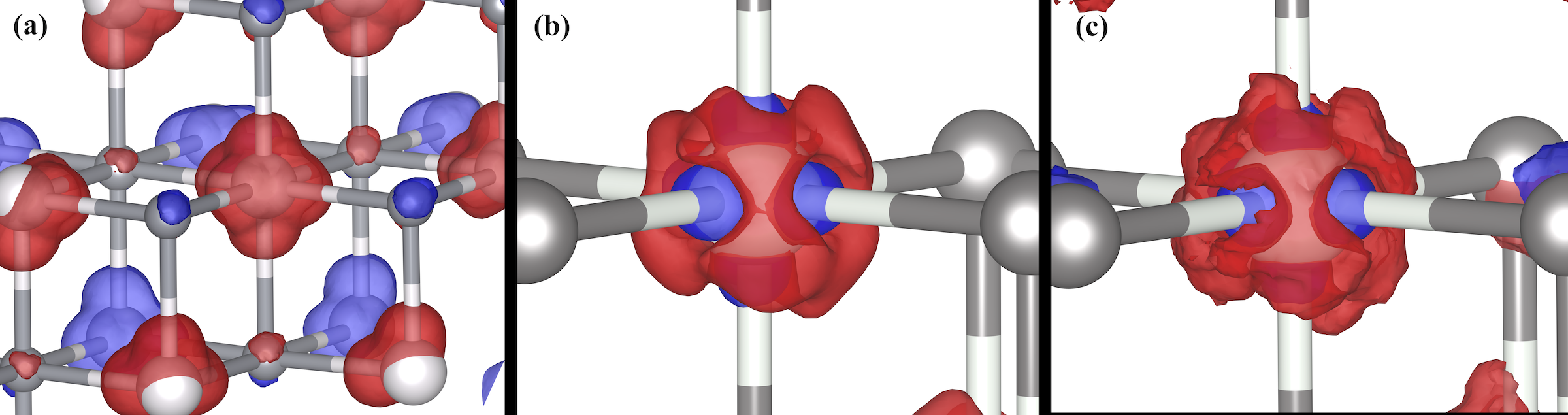}
\caption{(a): The DMC spin density in NiO with Ni atoms shown in white and O atoms shown in grey. Red (blue) represents positive (negative) spin densities in the cell. (b): The difference between the DMC and PBE0 spin density on a Ni site. Blue shows areas where DMC has a decreased spin density with respect to PBE0 and red shows areas where DMC increases the spin density. We see that DMC spreads spin density away from its bonds, explaining the decrease in its form factor at high values of \textbf{q} with respect to PBE0. (c): The difference between the DMC and PBE spin densities. PBE and PBE0 perform similarly in distributing the spin density despite the difference in the magnetic moments between the theories. All isosurfaces pictured are 0.03 $\mu_B$/$\AA^3$.}
\label{fig:spin}
\end{figure*}

Figure~\ref{fig:spin}a shows the DMC spin density of a Ni and its surrounding O environment.
In Figure~\ref{fig:spin}b, we have taken the difference between the spin density of DMC and PBE0 on a Ni site; despite these two theories yielding similar magnetic moments for Ni, there is a substantial difference in the organization of the spin on the transition metal site.
The blue (red) represents an area where DMC has removed (added) spin with respect to PBE0.
Comparing the two theories, DMC spreads the spin density away from the bond directions and smears it on the Ni site while PBE0 keeps the spin density along the bond directions.
Figure~\ref{fig:spin}c demonstrates similar behavior for the difference between DMC and PBE, although the extent of the differences is greater. 
This result is sensible from the perspective of the form factor since PBE0 and PBE have higher values of the form factor when compared to DMC at values of \textbf{q} associated with on-site correlations.

\section{Conclusion}
We have studied the use of DFT and DMC as a means of obtaining the experimental magnetic neutron intensities and form factors.
We calculate the structure factor in DMC and DFT and show agreement to about 10 percent for MnO.
In the case of NiO, both DFT functionals underestimate the structure factor with respect to DMC.
To ground our benchmark, we performed powder neutron diffraction experiments at MURR to determine the experimental intensity and form factor for both materials.
By calculating $S(\textbf{q)}$ and using the experimental magnetic moment direction, we produced theoretical intensities.
We show DMC predicts magnetic neutron intensities and form factors with higher accuracy than either PBE or PBE0 in both MnO and NiO.
While all the techniques accurately compute the intensity for MnO, we found that NiO requires the use of a hybrid method (PBE0) or an explicitly correlated method (DMC) to qualitatively compute the neutron intensity.
The error in PBE is primarily due to the underestimate of the magnetic moment on the Ni sites.

We further related the differences in $S(\textbf{q})$ to the real-space spin density.
We found both PBE and PBE0 have more spin density along the bond directions of NiO compared to DMC, which has a more isotropic shape. 
The magnetic form factor corroborates what we found in the spin density.
While DMC accurately calculates the form factor in NiO, PBE and PBE0 were shown to overestimate the atomic form factor, indicating the experimental spin density is less localized than that computed in DFT.
This information found herein could be used as a criterion for improved density functionals that can properly describe spin density in correlated materials.

\section{Acknowledgements} 
A.M. and L.K.W. were supported by a grant from the Simons Foundation as part of the Simons Collaboration on the many-electron problem.
Diffraction work (L.K., K.L, G.J.M.) was sponsored by the National Science Foundation, under grant number DMR-1455264-CAR. 
X-ray diffraction measurements were carried out in the Materials Research Laboratory Central Research Facilities at the University of Illinois.
\bibliography{munoz_neutron_benchmark}

\end{document}